\documentclass[lettersize,journal]{IEEEtran}
\IEEEoverridecommandlockouts
\usepackage{cite}
\usepackage{amsmath,amssymb,amsfonts}
\usepackage{algorithmic}
\usepackage{graphicx}
\usepackage{textcomp}
\usepackage{xcolor}
\def\BibTeX{{\rm B\kern-.05em{\sc i\kern-.025em b}\kern-.08em
    T\kern-.1667em\lower.7ex\hbox{E}\kern-.125emX}}

\usepackage{tabularx}
\usepackage{fancyhdr} % 添加页眉页脚
\usepackage{booktabs,multirow,tabularx}

\usepackage{setspace}
\begin{document}
	
\title{ISAC Prototype System for Multi-Domain Cooperative Communication Networks
}

\author{Jie Yang, \textit{Member, IEEE}, Hang Que, Tao Du, Le Liang, \textit{Member, IEEE}, Xiao Li, \textit{Member, IEEE}, \\ Chao-Kai Wen, \textit{Fellow, IEEE}, and Shi Jin, \textit{Fellow, IEEE}
		\thanks{Manuscript received 9 August 2024; accepted 28 October 2024. This work was supported in part by the National Natural Science Foundation of China (NSFC) under Grant 62301156 and 62341107, in part by the Fundamental Research Funds for the Central Universities 2242022k60004, 
			in part by the Key Technologies R\&D Program of Jiangsu (Prospective and Key Technologies for Industry) under Grants BE2023022 and BE2023022-1, and
			in part by the National Science Foundation of Jiangsu Province under Grant BK20230818. The work of C.-K. Wen was supported in part by the National Science and Technology Council of Taiwan under the grant MOST 111-2221-E-110-020-MY3 and by the Sixth Generation Communication and Sensing Research Center, which is funded by the Higher Education SPROUT Project of the Ministry of Education (MOE) of Taiwan. The associate editor coordinating the review of this article and approving it for publication was G. Chen. \textit{(Corresponding author: Shi Jin.)}}
		\thanks{ Jie Yang is with the Key Laboratory of Measurement and Control of Complex Systems of Engineering, Ministry of Education, Southeast University, Nanjing 210096, China (e-mail: yangjie@seu.edu.cn).
			
			Hang Que, Tao Du, Le Liang, Xiao Li, and Shi Jin are with the National Mobile Communications Research
			Laboratory, Southeast University, Nanjing 210096, China.
			
			Jie Yang, Le Liang, and Shi Jin are with the Frontiers Science Center for Mobile Information Communication and Security, Southeast University, Nanjing 210096, China.
			
			Chao-Kai Wen is with the Institute of Communications Engineering, National Sun Yat-sen University, Kaohsiung 80424, Taiwan. }	  	   		
				\vspace{-4mm}}
\maketitle

\begin{abstract}
Future wireless networks are poised to transform into integrated sensing and communication (ISAC) networks, unlocking groundbreaking services such as digital twinning. To harness the full potential of ISAC networks, it is essential to experimentally validate their sensing capabilities and the role of sensing in boosting communication. However, current prototype systems fall short in supporting multiple sensing functions or validating sensing-assisted communication. In response, we have developed an advanced ISAC prototype system that incorporates monostatic, bistatic, and network sensing modes. This system supports multimodal data collection and synchronization, ensuring comprehensive experimental validation. On the communication front, it excels in sensing-aided beam tracking and real-time high-definition video transmission. For sensing applications, it provides precise angle and range measurements, real-time angle-range imaging, and radio-based simultaneous localization and mapping (SLAM). Our prototype aligns with the 5G New Radio standard, offering scalability for up to 16 user equipments (UEs) in uplink transmission and 10 UEs in downlink transmission. Real-world tests showcase the system's superior accuracy, with root mean square errors of 2.3 degrees for angle estimation and 0.3 meters (m) for range estimation. Additionally, the estimation errors for multimodal-aided real-time radio SLAM localization and mapping are 0.25 m and 0.8 m, respectively.
\end{abstract}


\begin{figure*}[htbp]
	\centerline{\includegraphics[width=1\textwidth]{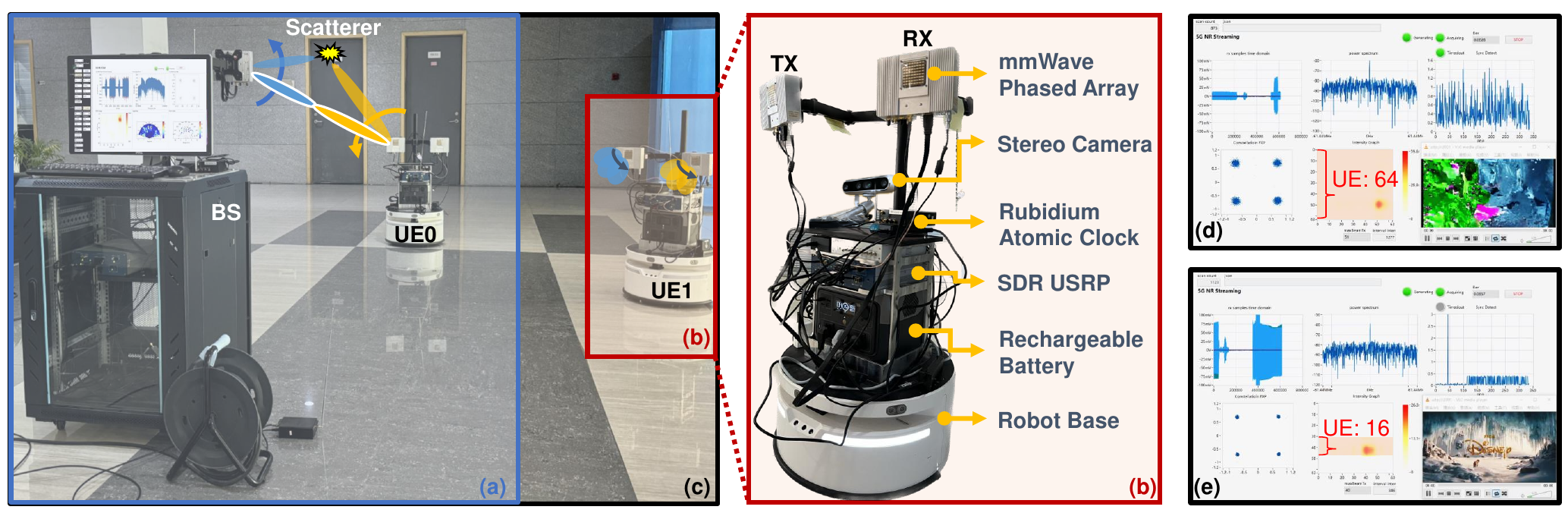}}
	\vspace{-1mm}
	\caption{Hardware architecture and front-end interface display. (a) Bistatic sensing. (b) Monostatic sensing. (c) Network sensing. (d) Full beam scanning. (e) Sensing-aided beam tracking. Sub-figures in (d) and (e) include time-domain waveform, power spectrum, synchronization head detection, constellation diagram, reference signal received power (RSRP) image, and real-time video receiving. Rows and columns of the RSRP image represent the beam indexes of the UE and BS, respectively. BS scans 64 beams, while UE searches 64 beams in full beam scanning and 16 beams in sensing-aided beam tracking.}
	\label{fig:Hard}
\end{figure*}

\section{Introduction}
\IEEEPARstart{F}{uture} wireless networks aim to achieve ubiquitous connectivity with ultra-high throughput, reliability, and ultra-low latency, while also enabling the ability to sense, control, and optimize wireless environments. Integrated sensing and communications (ISAC) seamlessly combine sensing and communication functionalities within a single system, enabling environment-aware wireless communication and promising significant advantages for future networks \cite{3GPP1,LF}. Recognized by the International Telecommunication Union as one of the six key application scenarios for 6G, ISAC represents a critical advancement. To fully harness ISAC's potential in wireless networks, it is essential to move beyond theoretical and simulation studies and validate these capabilities experimentally, assessing how sensing can enhance communication.

Several prototype systems have been developed to verify the potential benefits of integrating sensing and communication. For instance, sensing techniques like radio-based simultaneous localization and mapping (SLAM) have shown great promise by collecting data for offline validation using millimeter-wave (mmWave) \cite{RadioSLAM,RadioSLAM1} and Terahertz \cite{RadioSLAM2} communication systems. Prototyping and experimental results for environment-aware communications using the channel knowledge map (CKM) approach have demonstrated that CKM-based beam alignment can significantly reduce beam training overhead compared to exhaustive searches \cite{CKM}. Additionally, experiments in \cite{RadioSensing} confirmed the feasibility of achieving real-time environmental awareness while maintaining communication rates. However, these experiments are often limited to a single node or mode of operation.
Cooperative bistatic sensing is realized in \cite{li2024WCNC} using Doppler measurements, and cooperative monostatic sensing is explored in \cite{Chen2024JSAS} to achieve localization (meter-level accuracy), human activity recognition, and human skeleton imaging.
Network collaboration and multimodal fusion are emerging as future trends in ISAC development \cite{Cooperative,YJ}. 
Network sensing in system implementation must address the following practical issues: modeling sensing reliability for node selection and BS switching; establishing a unified fusion reference for varying scales and coordinate systems; reducing computational overhead for fusing large volumes of node data; and semantically compressing feedback sensing information.
Multi-domain cooperation in wireless networks involves cross-mode strategies such as monostatic sensing, bistatic sensing, and network sensing, alongside cross-frequency and cross-modal integration. Despite the promise of these approaches, there remains a shortage of prototype systems that enable multi-domain cooperation. 

In this paper, we address this gap by integrating monostatic, bistatic, and network sensing modes into a prototype system for the verification of multi-domain cooperative ISAC and sensing-aided communication \cite{QH}, in line with 5G New Radio (NR) standards \cite{3GPP2}. Notably, this system achieves real-time radio SLAM and SLAM-aided beam tracking for the first time. The contributions of this letter are twofold: 

First, we establish an ISAC prototype system where software-defined radio (SDR) devices at the transmitter and receiver integrate field-programmable gate array (FPGA) parallel computing chips and x86 processors. This integration enables rapid baseband signal processing and scalable sensing applications. The prototype system utilizes an OFDM waveform for communication and sensing, operating at 28 GHz with an effective bandwidth of 95 MHz. It supports a maximum instantaneous downlink rate of 508 Mbps and provides plug-and-play functionality for multimodal sensor data, including stereo cameras and inertial measurement units (IMU). The system features monostatic sensing for real-time angle-range imaging, bistatic sensing for real-time radio SLAM, and network sensing for multi-domain cooperation.

Second, the experimental results validate the system's sensing capabilities and demonstrate the necessity of sensing-aided communications. Specifically, multimodal calibration effectively mitigates dynamic interferences and antenna array jitter in real-time radio SLAM, achieving localization accuracy of 0.25 meters (m) and mapping accuracy of 0.8 m. Exhaustive beam search under 5G NR standards during user movement can cause beam misalignment and communication interruptions. Leveraging the system's sensing capabilities, sensing-aided beam tracking significantly reduces beam search overhead and resolves beam misalignment issues due to mobility.

\begin{table}
	\centering
	\footnotesize
	\caption{Main System Parameters}
	\vspace{-0mm}
	\label{parameters}
	\begin{tabular}{l  l  l  l }
		\toprule[1.5pt]
		\bfseries{Parameter} & \bfseries {Value} &\bfseries{Parameter} & \bfseries {Value} \\		
		\midrule		
		Carrier Frequency & $28$ GHz & Antenna Array & $8\times 8$\\
		ISAC Waveform & OFDM & Sampling Rate& $122.88$ MHz\\
		Effective Subcarriers & $792$ & Subcarrier Spacing & $120$ KHz \\
		Effective Bandwidth & $95$ MHz & Transmission Rate & $508$ Mbps\\		
		Angle RMSE & $2.3^{\circ}$ & Range RMSE & $0.3$ m\\ 
		\bottomrule[1.5pt]
	\end{tabular}
\end{table}

\begin{figure}
	\centerline{\includegraphics[width=0.499\textwidth]{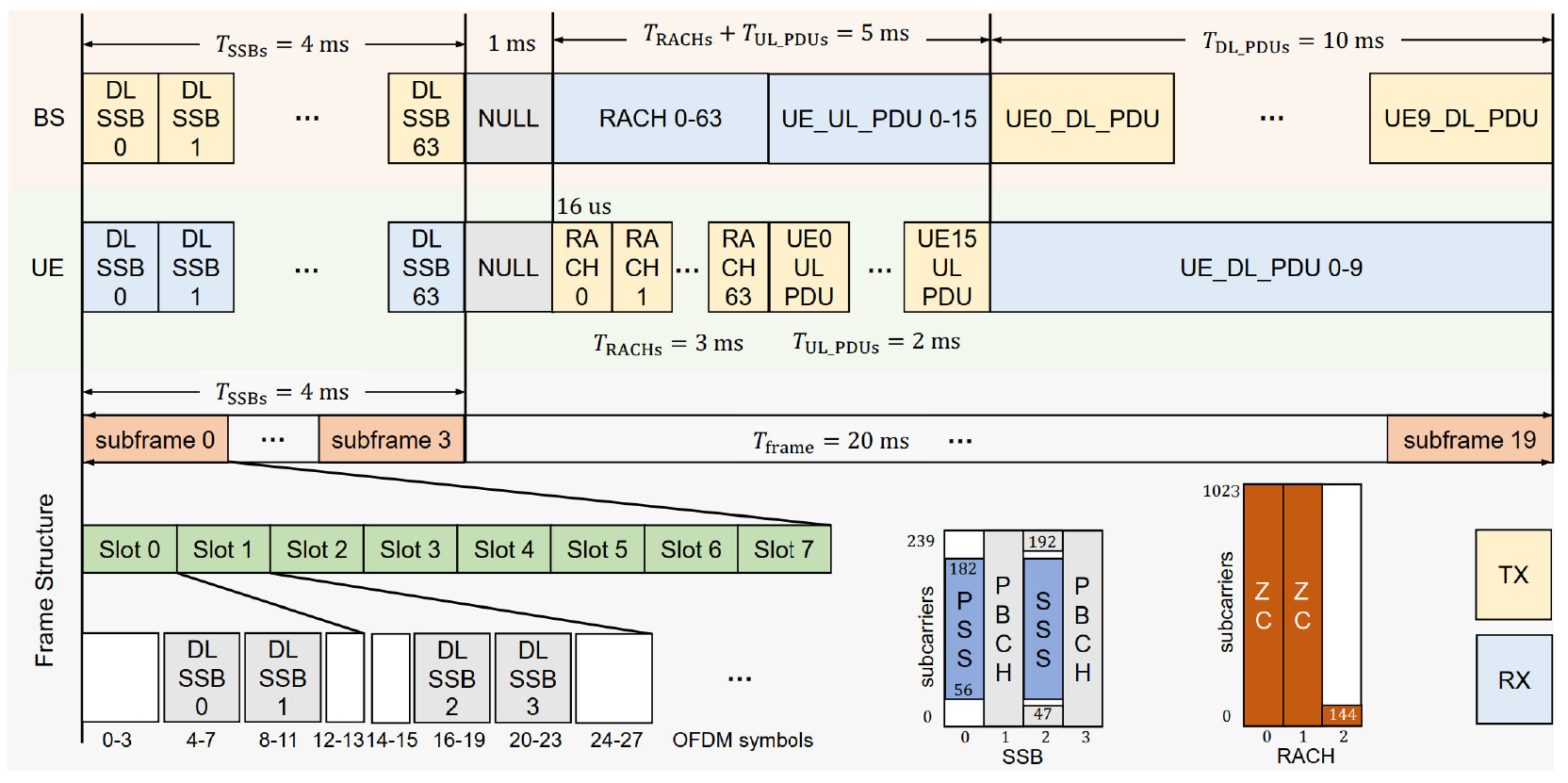}}
	\vspace{-0mm}
	\caption{Frame structure that support multi-nodes cooperation. DL: downlink, UL: uplink, SSB: synchronization signal block, RACH: random access channel, PDU: protocol data unit, PSS: primary synchronization signal, PBCH: physical broadcast channel, SSS: secondary synchronization signal, ZC: Zadoff-Chu, TX: transmit, RX: receive.}
	\label{fig:Frame}
	\vspace{-0mm}
\end{figure}

\section{System Overview}
The proposed ISAC prototype system for multi-domain cooperative communication networks is briefly illustrated in Fig. \ref{fig:Hard}, and the applications are presented as follows:

\begin{itemize}
    \item \textbf{Bistatic sensing} [Fig. \ref{fig:Hard}(a)]: The base station (BS) performs downlink beam scanning while the user equipment (UE) switches beams for reception, enabling a bistatic sensing mode where the BS transmits and the UE receives. This principle also applies to the uplink.

    \item \textbf{Monostatic sensing} [Fig. \ref{fig:Hard}(b)]: The node is equipped with two mmWave phased arrays, one for transmission and the other for reception, enabling a self-transmitting and self-receiving monostatic sensing mode.

    \item \textbf{Network sensing} [Fig. \ref{fig:Hard}(c)]: The BS simultaneously supports multiple UEs, with each UE using either bistatic or monostatic sensing to monitor their local radio environments. The cooperative UEs then transmit their sensing results to the BS for potential fusion. This collaboration enhances sensing accuracy and expands the sensing range. The same principle applies to multiple BS scenarios. 
        
    \item \textbf{Communication} [Fig. \ref{fig:Hard}(e)]: The BS and UE follow 5G NR standards and utilize the matched beam pair for downlink and uplink communications. The system supports multiple modulation schemes from 4 quadrature amplitude modulation (QAM) up to 64 QAM. Real-time downlink video transmission validates the sensing-aided beam tracking scheme.
\end{itemize}

The proposed ISAC prototype system possesses the following advantages:
\begin{itemize}
    \item \textbf{Multifunctional}: Realizes real-time video transmission, angle, range, Doppler measurement, localization, and environmental reconstruction (e.g., imaging and SLAM) capabilities.

    \item \textbf{Standardized}: Designed according to the 5G NR standard with the OFDM waveform. Sensing functions are realized during pilot and data transmission processes to facilitate communication-centric ISAC.

    \item \textbf{Scalability}: Supports distributed nodes, facilitating the validation of theories and technologies of network collaboration in ISAC.

    \item \textbf{Interoperability}: Facilitates seamless communication and collaboration among diverse domains, including different sensing modes and devices. Devices are synchronized and have plug-and-play capabilities to enable testing of multi-modal data fusion.
\end{itemize}

\section{Prototype Setup}

In this section, we describe the hardware architecture and the frame structure of the proposed ISAC prototype system. The main system parameters are summarized in Table \ref{parameters}.

\subsection{Hardware Architecture}

The hardware architecture of the ISAC prototype system (Fig. \ref{fig:Hard}) primarily comprises the SDR National Instruments USRP2974, the mmWave phased array mmPSA-TR64MX, the rubidium atomic clock SYN3306, the robot base WATER2, the stereo camera Intel RealSense D455, and a 220V rechargeable battery. The SDR supports a bandwidth of up to 160 MHz and a frequency range spanning from 10 MHz to 6 GHz. The intermediate frequency output of the SDR is 2.5 GHz, which is then up-converted to 28 GHz radio frequency (RF) by the local oscillator built into the mmWave phased array. The mmWave phased array is a uniform plane array with $8\times 8$ elements and a single RF chain, capable of achieving two-dimensional dynamic beam scanning with up to 4096 beam states within a horizontal and vertical range of $\pm 60$ degrees. Each node is equipped with two antenna arrays.

The rubidium atomic clock is essential for indoor reception of outdoor global positioning system signals, providing a synchronized timekeeping baseline for multiple nodes.
The robot base controls the movement of the UE along predefined trajectories, orientations, and velocities, offering the ground truth reference and IMU measurements at a frequency of 1 Hz. The stereo camera outputs depth images and RGB images, providing plug-and-play visual measurements. The stereo camera has a data acquisition frequency of 15-60 Hz. The front-end interface of the system is shown in Figs. \ref{fig:Hard}(d) and (e), which displays a wealth of system parameters and measured results, including RSRP image, angle-delay image, and point cloud image.

%The rubidium atomic clock is capable of outputting sine waves or square waves at 1PPS/XPPS/1MHz/5MHz/10MHz frequencies, and the phase noise can be as low as -150dBc/10kHz.

%, the field of view (FOV) range of depth is 87 (H)×58 (W), the FOV range of RGB image FOV range is 90 (H)×65 (V), a depth range of [0.6, 6] m.%, a maximum resolution of 1280×720.

\begin{figure}
	\vspace{-3mm}
	\centerline{\includegraphics[width=0.40\textwidth]{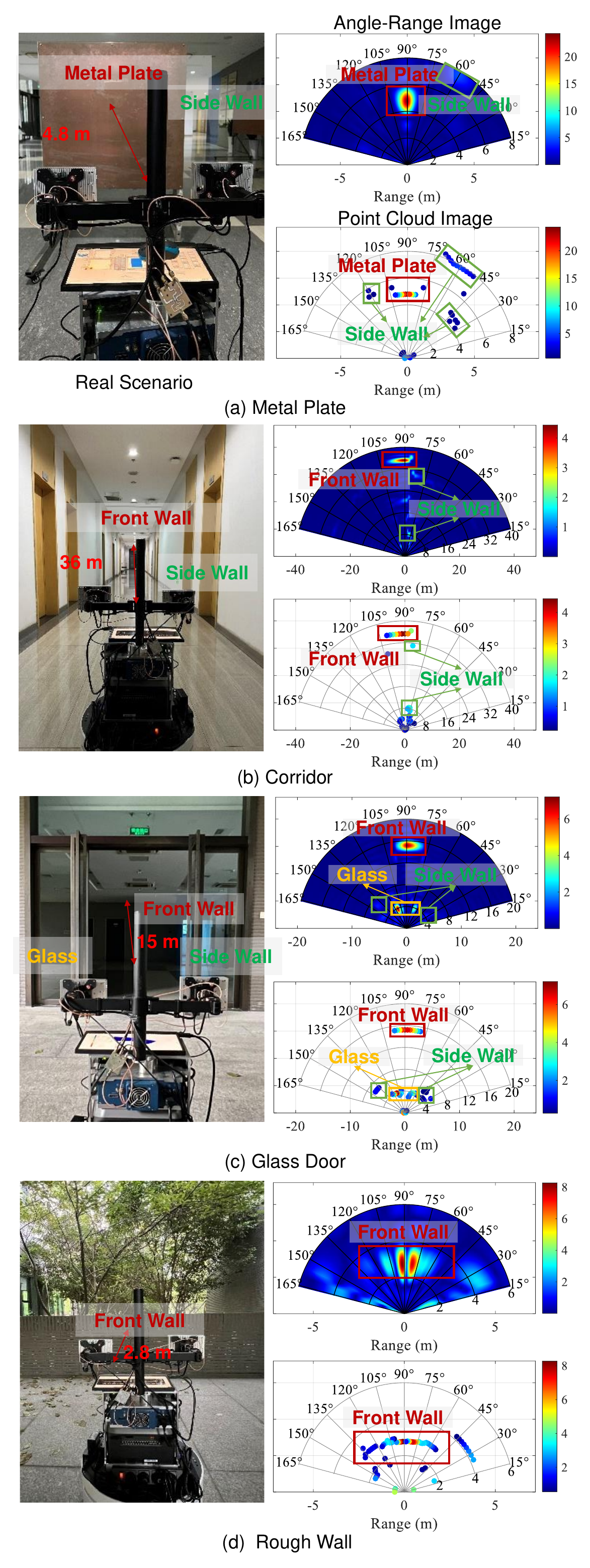}}
	\vspace{-3mm}
	\caption{Angle-range imaging and the corresponding point cloud.}
	\label{fig:imaging}
\end{figure}

\subsection{Frame Structure}

The frame structure is designed in accordance with the 5G NR standard (Fig. \ref{fig:Frame}). The length of a radio frame is 20 milliseconds (ms), consisting of 20 subframes. Each subframe includes 8 slots, with every two adjacent time slots forming a group, totaling 28 OFDM symbols. The number of effective subcarriers is 792, the subcarrier spacing is 120 kHz, and the corresponding effective bandwidth and sampling rate are 95.04 MHz and 122.88 MHz, respectively. As depicted in Fig. 2, a synchronization signal block (SSB) consists of four OFDM symbols in time and 240 subcarriers in frequency, and includes a primary synchronization signal (PSS), a secondary synchronization signal (SSS), and a physical broadcast channel (PBCH). The PBCH can be used to estimate the RSRP of the SSB. The random access channel (RACH) uses a Zadoff-Chu sequence of length 139, which generates 64 preambles through cyclic shifting. After undergoing fast Fourier transformation, zero padding, cyclic prefix insertion, upsampling, and other preprocessing steps, a frequency domain RACH signal of length 2,192 is generated, occupying less than three OFDM symbols. The frame structure supports up to 16 UEs for uplink transmission and 10 UEs for downlink transmission.

For the beam tracking process, the BS continues to broadcast 64 SSBs in the first 4 ms of each frame, corresponding to 64 beam indexes. UEs use the results from sensing algorithms to reduce the beam search space. 
Assume there are $M$ beams to search, denoted as B$_1$ to B$_M$, where $ M \ll 64$. The UE sequentially uses each beam for reception. Specifically, it uses B$_1$ for 4 ms, then after 20 ms, switches to B$_2$ for another 4 ms, and continues this pattern until all beam pairs have been traversed. Based on the index of the peak value in $M \times 64$ RSRP measurements, the UE identifies the beam pair indexes. The total beam tracking time is $20M$ ms. Therefore, the overhead of beam scanning is reduced by $(64-M)/64$. $M$ can be flexibly adjusted to accommodate various environments and movement speeds in our prototype system.

\begin{figure*}
	\vspace{-0mm}
	\centerline{\includegraphics[width=1\textwidth]{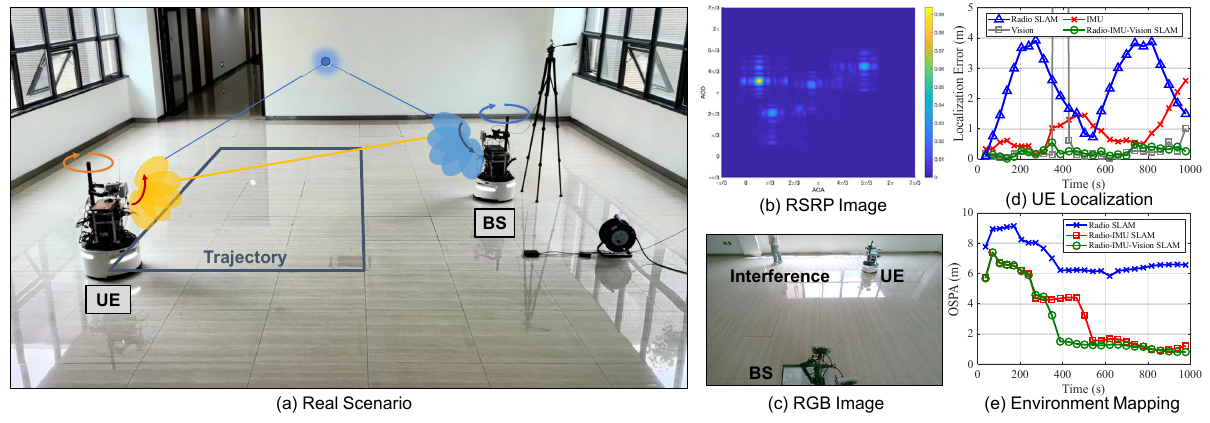}}
	\vspace{-0mm}
	\caption{Multimodal-aided radio SLAM.}\vspace{-1mm}
	\label{fig:SLAM}	\vspace{-1mm}
\end{figure*} 

\section{Experiments Results}

\subsection{Angle-Range Imaging}
Angle measurement is necessary in the bistatic sensing mode, as we typically do not know the angles of departure (AODs) and angles of arrival (AOAs). Our system can generate a $64 \times 64$ RSRP matrix after completing the $64 \times 64$ beam scanning (one BS-UE orientation pair). By combining RSRP matrices from different BS-UE orientation pairs, we can compile an RSRP image with extended angular coverage. The RSRP image represents the received signal in the beam space, which motivates us to transform the AOA and AOD estimation problem in radio frequency into an object detection problem in images. We propose a deep learning-based framework for estimating, tracking, and identifying mmWave channel multipath parameters as well as distinguishing between LOS and NLOS paths, transforming the entire process into a universal computer vision problem \cite{CK}. The real measurement results indicate that the root mean square error (RMSE) of angle estimation is $2.3^{\circ}$.

Range measurement is performed in the monostatic sensing mode, which eliminates the range measurement errors caused by clock asynchronization in bistatic sensing. In the monostatic sensing mode, we align the beamforming of the transmitting and receiving arrays in the same direction, enabling the receiving array to capture echo signals from scatterers. The relative positions of the transmitting and receiving arrays are fixed in our system, and self-interference typically occurs in a fixed area in the beam space. As a result, self-interference can be easily filtered out by exploiting its spatial orthogonality with the targets. Real measurement results show that the RMSE for range estimation in single-target scenarios is 0.3 m.
 
An angle-range image can be obtained in the monostatic sensing mode when the transmitting and receiving arrays switch to the same beam index simultaneously and traverse 64 beam directions. The monostatic sensing mode is completed during the random access process without competing for communication resources. 
Specifically, we form Fig. \ref{fig:imaging} by obtaining the power delay profile (PDP) for each beam direction. Then, a point cloud image is obtained by recording the position of the maximum peak point of the PDP in each direction, where the color of the target points represents the energy of the corresponding echo signals. The real measurements in Fig. \ref{fig:imaging} include imaging of a metal plate, a corridor, a glass door, and a rough wall.
The angle-range image obtained by our system shows angle and range spread. Firstly, the range spread is mainly determined by the system bandwidth. Secondly, the angle spread partly results from the beam having a certain width and partly reflects the shape of the target. In addition, targets closer to the normal direction of the array can be observed and estimated more accurately.

\vspace{-0mm}
\subsection{Multimodal-Aided Radio SLAM}

Real-time Radio SLAM with angle measurements \cite{RadioSLAM} is achieved using the bistatic sensing mode. The bistatic sensing mode operates entirely during normal communication processes, without consuming additional time-frequency resources, and only requires extra computational resources for processing sensing algorithms. 
The actual measurement scenario is shown in Fig. \ref{fig:SLAM}(a), where the blue-gray line represents the UE's trajectory. The UE moves continuously along the trajectory, stopping automatically at designated points and rotating in four orientations to obtain RSRP measurements covering $360^{\circ}$. The entire measurement process for one trajectory takes 20 minutes, during which the UE moves and measures autonomously without human intervention. However, continuous real-time data collection over extended periods faces challenges such as dynamic interferences (e.g., pedestrians), complex scatterers (e.g., non-specular reflections), and array surface jitter caused by the movement of equipment in the environment.

Inspired by the fact that human perception relies on multiple modalities---such as touch, sight, hearing, smell, and taste---even when some sensory inputs are unreliable, humans can still extract useful information from imperfect multimodal signals to piece together a coherent understanding of their surroundings. Similarly, we designed the multimodal-aided radio SLAM. 
In Radio-IMU-Vision SLAM, we first obtain the positional coordinates of targets within the field of view (FOV) using stereo camera-based computer vision tools. Notably, visual sensing detects multiple targets without distinguishing which one corresponds to the UE. By associating the UE's location from Radio-IMU SLAM with the visual results, we can accurately determine the UE's position within the visual outputs, enabling effective fusion of location data from both modalities at the decision level.

We can evaluate the strengths and weaknesses of different modalities by comparing the performance of multimodal-aided radio SLAM. Fig. \ref{fig:SLAM}(d) shows localization results for various single-modal measurements alongside multimodal-aided radio SLAM. The red line represents IMU localization, which deteriorates over time due to cumulative errors, resulting in significant drift after 800 seconds (s). The IMU provides relatively accurate short-term displacement and orientation estimates, but long-term accuracy degrades. The blue line shows vision-based localization, which is highly sensitive to lighting conditions, with accuracy decreasing at the edges of the FOV. Around 400 s, the target is lost due to floor reflections, and in simulations, a larger error is assigned when this occurs. The green line represents multimodal-aided radio SLAM, which performs the best with an average localization error of 0.25 m. The fixed position of the BS serves as a natural landmark, providing a strong basis for correction in self-localization systems like IMUs.
Fig. \ref{fig:SLAM}(e) compares the mapping performance of Radio SLAM, Radio-IMU SLAM, and Radio-IMU-Vision SLAM. The green line shows that Radio-IMU-Vision SLAM achieves an average mapping accuracy of 0.8 m, outperforming Radio SLAM and demonstrating faster convergence than Radio-IMU SLAM.

Finally, we compared the average iteration time of different SLAM algorithms on our host computer configuration (GPU: 4070ti + CPU: 13700F). Radio-IMU-Vision SLAM takes 0.023 s per iteration, while both Radio-IMU SLAM and Radio SLAM take 0.008 s. Despite the additional computational overhead introduced by multimodal methods, the system still meets real-time SLAM requirements. Furthermore, the improvements in accuracy and robustness justify the increased computational cost.

\subsection{Sensing-Aided Beam Tracking}
\vspace{-1mm}

We evaluate communication quality and beam tracking effectiveness visually through real-time high-definition video transmission. 
The mobile UE estimates its current position based on the previous position output by radio SLAM, combined with the current orientation and velocity information from the IMU on the robot base. The UE then predicts the incoming beam directions of both the LOS and specular reflection NLOS paths using the wireless environment information (including the locations of the BS and major reflecting surfaces) constructed by radio SLAM. Due to the estimation error in radio SLAM, the optimal beam index is typically not the predicted value but one close to it. In our system, the UE can flexibly select $M \in [1,64]$ beams adjacent to the predicted beam direction for small-range beam scanning. Finally, the highest quality path is selected from the candidate multiple paths for communication. The rows and columns of the RSRP image [the fifth sub-figure in Fig. \ref{fig:Hard}(d)] represent the beam indexes of the UE and BS, respectively.
When the sensing-aided beam tracking function is enabled, the rows of $M$ beams in the RSRP image are activated [the fifth sub-figure in Fig. \ref{fig:Hard}(e)], while the remaining rows have no data.

In the experiments, we assess the impact of different beam scanning ranges on communication performance. The UE moves linearly at a speed of 4 meters/second (m/s), receiving a video segment for 200 s transmitted from the BS.
The results show that when the UE performs exhaustive beam searching, video reception was interrupted five times, totaling 47 s of interruptions [Fig. \ref{fig:Hard}(d)].	
Due to the movement of the UE, the rate of optimal beam changes is faster than the completion of a full exhaustive search (1.28 s), causing beam misalignment and degraded communication quality. Sensing-aided beam tracking reduces the beam scan range from 64 to 16 beams on the UE side. As a result, the RSRP image size changes from $64 \times 64$ [Fig. \ref{fig:Hard}(d)] to $16 \times 64$ [Fig. \ref{fig:Hard}(e)], reducing the scan time to 0.32 s.	
Only one interruption occurred, demonstrating that sensing-assisted beam tracking enables high-definition video transmission and resolves beam misalignment issues in dynamic scenes---something exhaustive searches could not achieve---while reducing beam search overhead by 75\%.

\section{Conclusion}
We propose an advanced ISAC prototype system that incorporates monostatic, bistatic, and network sensing modes, enabling multimodal data collection and synchronization for comprehensive experimental validation. Experiments confirm the system's capabilities in real-time angle-range imaging and radio SLAM. Additionally, the system supports high-definition video transmission through sensing-aided beam tracking, effectively addressing beam misalignment issues in dynamic scenes. Future research will leverage this system to explore network collaboration, aiming to expand the sensing range and enhance the accuracy of individual nodes.

% Comment: 參考論文格式再檢查一次
% 1) 作者三人以上，可以用 et al. 
% 正確的用法應該是 F. Liu \textit{et al.}
% 不正確: J. Yang, C. -K. Wen, J. Xu, \textit{et al.} 
% 格式要一致
% 2) 會議論文 按標準的規範來
% 3) arXiv 這些，更新論文狀態 ，看是否登出

\end{document}